\documentclass[12pt]{article}

\usepackage{arxiv}

\usepackage[utf8]{inputenc} % allow utf-8 input
\usepackage[T1]{fontenc}    % use 8-bit T1 fonts
\usepackage[hidelinks]{hyperref}       % hyperlinks
\usepackage{url}            % simple URL typesetting
\usepackage{booktabs}       % professional-quality tables
\usepackage{amsfonts}       % blackboard math symbols
\usepackage{nicefrac}       % compact symbols for 1/2, etc.
\usepackage{microtype}      % microtypography
\usepackage{graphicx}
\usepackage{epstopdf, epsfig}
\usepackage{amsmath}
\usepackage{graphicx}
\usepackage{graphics}
\usepackage{listings}

\usepackage{natbib}	% bibliography style
\usepackage{gensymb}
\usepackage{comment}
\usepackage{enumitem}
\setitemize{noitemsep,topsep=0pt,parsep=0pt,partopsep=0pt,leftmargin=*}
\usepackage{amssymb}

\usepackage{lineno}
\usepackage{setspace}
\usepackage{footnote}

\usepackage{tabularx}

\usepackage{nopageno}

\usepackage{fancyhdr}
\usepackage{subcaption}

\usepackage{xcolor}
\usepackage{multirow}

\title{Grid turbulence measurements with an acoustic Doppler current profiler }
\author{
  Trygve. K. Løken$^{a,}$\footnotemark, David Roger Lande-Sudall$^{b}$, Atle Jensen$^{a}$ and Jean Rabault$^{c}$
}

\begin{document}
\maketitle
 
\footnotetext{E-mail address and phone number for corresponding: trygvekl@math.uio.no (+47) 94885767 (T.K. Løken)\\ 
E-mail addresses: atlej@math.uio.no (A. Jensen), jean.rblt@gmail.com (J. Rabault).}

\noindent $^{a}$ Department of Mathematics, University of Oslo, Oslo, Norway \\
$^{b}$ Western Norway University of Applied Sciences, Bergen, Norway \\
$^{c}$ Norwegian Meteorological Institute, Oslo, Norway \\

%\doublespacing
%\linenumbers

\begin{abstract}

The motivation for this study is to investigate the abilities and limitations of a Nortek Signature1000 acoustic Doppler current profiler (ADCP) regarding fine-scale turbulence measurements. Current profilers offer the advantage of gaining more coherent measurement data than available with point acoustic measurements, and it is desirable to exploit this property in laboratory and field applications. The ADCP was tested in a towing tank, where turbulence was generated from a grid towed under controlled conditions. Grid-induced turbulence is a well-studied phenomenon and a good approximation for isotropic turbulence. Several previous experiments are available for comparison and there are developed theories within the topic. In the present experiments, an acoustic Doppler velocimeter (ADV), which is an established instrument for turbulence measurements, was applied to validate the ADCP. It was found that the mean flow measured with the ADCP was accurate within 4\% of the ADV. The turbulent variance was reasonably well resolved by the ADCP when large grid bars were towed at a high speed, but largely overestimated for lower towing speed and smaller grid bars. The effective cutoff frequency and turbulent eddy size were characterized experimentally, which provides detailed guidelines for when the ADCP data can be trusted and will allow future experimentalists to decide \textit{a priori} if the Nortek Signature can be used in their setup. We conclude that the ADCP is not suitable for resolving turbulent spectra in a small-scale grid-induced flow due to the intrinsic Doppler noise and the low spatial and temporal sample resolution relative to the turbulent scales.

\end{abstract}

\section{Introduction} \label{sec:introduction}

Due to their long profile range, autonomy and simple principals of operation, acoustic Doppler current profilers (ADCPs) are frequently used for turbulence measurements in the ocean and tidal channels \citep{stacey1999measurements,guion2014frequency,guerra2017turbulence,shcherbina2018observing}. As the technological advances keep improving the spatial and temporal resolution of the instruments, small-scale turbulence in the ocean, such as wake formation behind floating ice or human structures, may be studied with ADCPs. Some authors have also reported on turbulence measurements with ADCPs in laboratory facilities \citep{hardcastle1997investigation,nystrom2007evaluation}. However, ADCPs have a measurement volume ($\sim100~\mathrm{cm}^{3}$) that is typically much greater than the smallest eddy structures of the flow. Although the smallest structures cannot be resolved, some quantities such as the turbulent kinetic energy (TKE) primarily depend on the large energetic eddies and may therefore be estimated with ADCPs \citep{nystrom2007evaluation}. Due to its random nature, turbulence is usually described through statistical parameters like TKE, variance, TKE spectra and Reynolds stresses \citep{nystrom2002measurement}. The parameters that can be reasonably well estimated are usually first and second-order statistical properties because Doppler noise places a constraint on the accuracy of the instantaneous velocity estimates \citep{mcmillan2016rates}.  

Various configurations with three to five beams are available for ADCPs \citep{dewey2007reynolds}. A five beam broadband Nortek Signature1000 (kHz) ADCP was utilized in this study. The instrument has one vertical oriented beam $\mathbf{b_5}$ and four slanted beams $\mathbf{b_1-b_4}$ diverging at $\theta=25^{\circ}$ from the vertical and separated at $90^{\circ}$ in the horizontal plane, i.e. in a Janus configuration. Five transducers emit acoustic waves along each beam, which are backscattered by particles suspended in the water. The particles are assumed to passively follow the fluid motion. The along-beam velocity component of the particles is calculated internally in the instrument from the Doppler shift of the reflected signal. Positive direction is defined radially away from the instrument and the beam velocities are denoted $b_j$ for $j=1,2,...,5$. Each beam is divided into several cells that can be as small as 2~cm when the instrument is operated in the pulse-to-pulse coherent mode, also known as the \textit{high-resolution} (HR) mode. Velocity profiles in three-dimensional space are therefore obtained, which increase the spatial distribution of data compared to traditional velocimeters, such as acoustic Doppler velocimeters (ADVs), that measure in a single point. This is a central motivation for trying to use ADCPs in laboratory applications. ADCPs perform non-intrusive measurements and the possibility of flow disturbance is practically non-existing over most of the profile \citep{nystrom2002measurement}, as long as the instrument axis $\mathbf{b_5}$ is perpendicular to the dominating flow direction, so that its wake does not pollute the measurement domain. Another advantage of ADCPs is that they require no calibration, just occasional maintenance and operation checks \citep{lipscomb1995quality}.   

However, ADCPs have limitations that must be kept in mind when configuring the instruments and processing the data. Acoustic velocimeters measure the phase shift $\phi$ of the backscattered signal, which lies in the range of $-180^{\circ}$ to $180^{\circ}$. If the particle velocity exceeds the velocity range associated with the instrument-specific ambiguity velocity $U_{amb}$, this will yield a corresponding phase shift outside the expected phase range, leading to ambiguity errors. The ambiguity velocity is defined as

\begin{equation}
\label{eq:U_amb}
U_{amb}=\frac{c}{8F_0\tau},
\end{equation}

\noindent where $c$ is the speed of sound in water, $F_0$ is the sonar carrier frequency and $\tau$ is the time-lag between two consecutive pulses \citep{shcherbina2018observing}. These errors can be identified as large steps in the time series and need to be corrected ("unwrapped") in the post processing. When the ADCP is operated in the HR mode, $U_{amb}$ is quite low, so there is a trade-off between the cell size and velocity range (see e.g. \cite{lhermitte1984pulse}). In addition, acoustic instruments have intrinsic Doppler noise $n$ in the beam velocity measurements, which is caused when the Doppler shift is estimated from finite-length pulses \citep{voulgaris1998evaluation}. Also, ringing and sidelobe interference may lead to errors close to the transducer and solid boundaries, respectively. Ringing is caused when the transducers continue to vibrate for a short time after the acoustic wave has been emitted, and the instrument cannot accurately record the backscattered signals until the transducer membranes have settled \citep{nystrom2007evaluation}. 

In many of the above-mentioned studies, ADV measurements are used as ground-truth values for comparison with ADCPs. ADVs are usually more accurate than ADCPs due to their small measurement volume ($\sim1~\mathrm{cm}^{3}$), lower Doppler noise and higher temporal resolution. For example, \cite{voulgaris1998evaluation} found that an ADV was able to reproduce turbulence properties (Reynolds stresses) in a laboratory facility within 1\% of a laser Doppler velocimeter. The present study is an extension of the previous work where the ability of an ADCP to measure grid-generated turbulence properties is investigated and comparisons are made with a Nortek Vectrino ADV. To our knowledge, turbulence from a towed grid has not yet been evaluated with an ADCP, hence this work is a new contribution to the literature.

Grid-induced turbulence is a phenomenon that has been studied in wind tunnels \citep{comte1966use,uberoi1967effect,comte1971simple,sirivat1983effect}, as well as in water tanks and flumes \citep{liu1995energetics,murzyn2005experimental}. In wind tunnels and water flumes, the grid is fixed in space and the fluid flows past it. In water tanks, the grid and instruments are usually towed along the axial direction of the tank, which puts a constraint on the duration of each repetition \citep{liu1995energetics}. Most studies related to grid-induced turbulence focus on the decay of turbulent properties, such as velocity variance and TKE, as function of the normalized downstream distance to the grid $x/M$, where $x$ is the downstream distance and $M$ is the mesh size. For example, \cite{sirivat1983effect} deduced a power law for TKE in the vertical direction $TK_z$, non-dimensionalized over the fluid speed $U_f$ squared, $TK_z/U_{f}^2=a(x/M)^{m}$, where $m\approx-1.3$ is independent of the fluid speed and $a$ is a constant. \cite{liu1995energetics} obtained the same power law and coefficient $m$ in a water tank, where the towing speed $U$ was the equivalent to the fluid speed. 

In grid-induced turbulence, structures coexist over a range of spatial scales $l$, where $l=2\pi/k$ and $k$ is the turbulent wavenumber. The integral scale $L$ corresponds to the largest turbulent eddies where TKE is produced, which are generated from grid-water interactions. Although the turbulence developing downstream of a grid is isotropic (directionally independent) and homogeneous (spatially independent) in theory, this is not obtained in reality at the integral scales \citep{comte1966use}. As TKE cascades to the increasingly smaller structures, the flow is assumed locally isotropic and the TKE wavenumber spectra should be proportional to $k^{-5/3}$ according to the Kolmogorov law for developed turbulence \citep{kolmogorov1941the}. This power law is valid in the inertial subrange that comprises scales $l_{IS}$, where $L \gg l_{IS} \gg l_{K}$ and $l_{K}$ is the Kolmogorov micro-scale at which TKE is dissipated into heat due to viscosity. Since temporal measurements are made in this study, the turbulent wavenumber is related to the eddy frequency $f$ through Taylor's hypothesis for steady state turbulence: $k=2\pi f/\langle u \rangle$, where $\langle u \rangle$ is the time averaged velocity at which the flow is advected past the instruments. Hence, the frequency spectra should be proportional to $f^{-5/3}$ in the inertial subrange.

The aim of the present study is to investigate the ability of the Nortek Signature, operated in the HR mode, to accurately resolve velocity variance and other TKE properties in fine-scale turbulence under well defined flow conditions. The turbulence was generated from a regular grid that was towed in a tank of quiescent fresh water. The paper is organized in the following manner. Section~\ref{sec:setup} describes the experimental facility and setup. The processing algorithms are given in Section~\ref{sec:processing} and the main findings of the study are presented and discussed in Section~\ref{sec:results}. Finally, the concluding remarks are summarized in Section~\ref{sec:conclusion}.

\section{Experimental setup} \label{sec:setup}

\begin{figure}
	\centering
	\subcaptionbox{Grid seen from downstream.\label{fig:towing_config_a}}
		{\includegraphics[width=.4\linewidth]{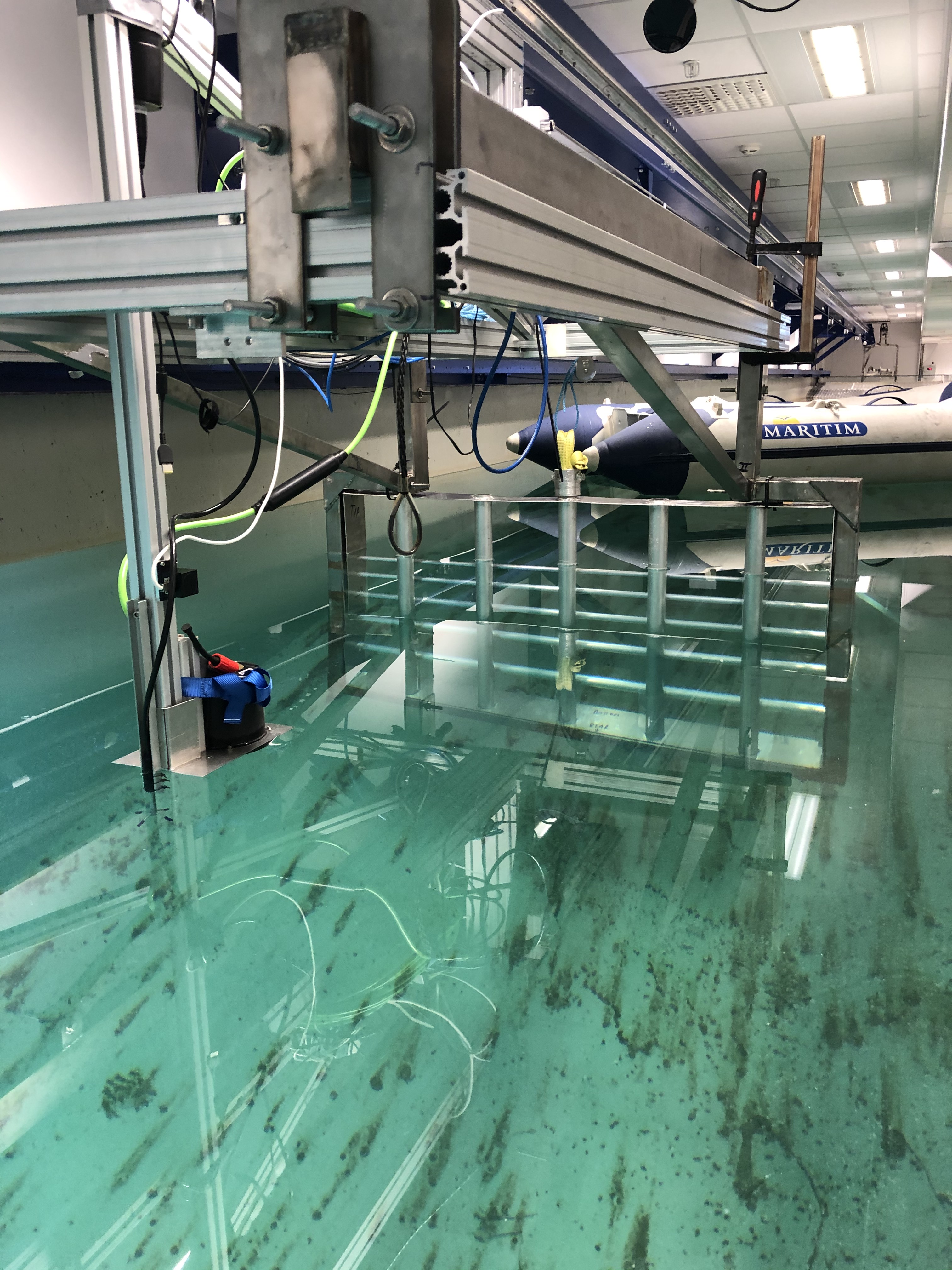}} 
	\subcaptionbox{Grid seen from upstream.\label{fig:towing_config_b}}
		{\includegraphics[width=.4\linewidth]{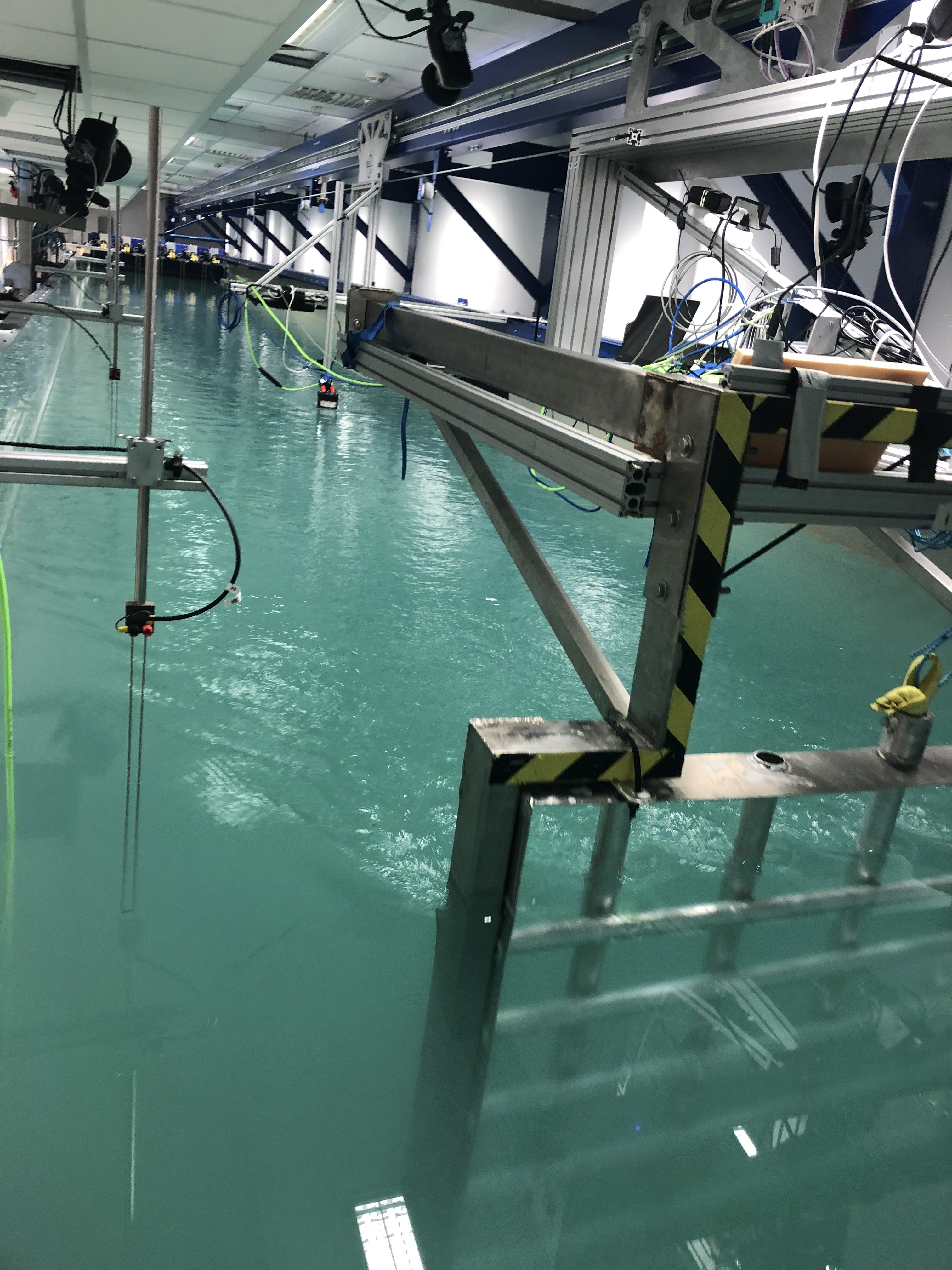}}
	\caption{ADCP suspended from the main carriage during standstill (a) and from the second carriage during towing (b).}\label{fig:towing_config}
\end{figure}

\begin{figure}
  \begin{center}
    \includegraphics[width=.8\textwidth]{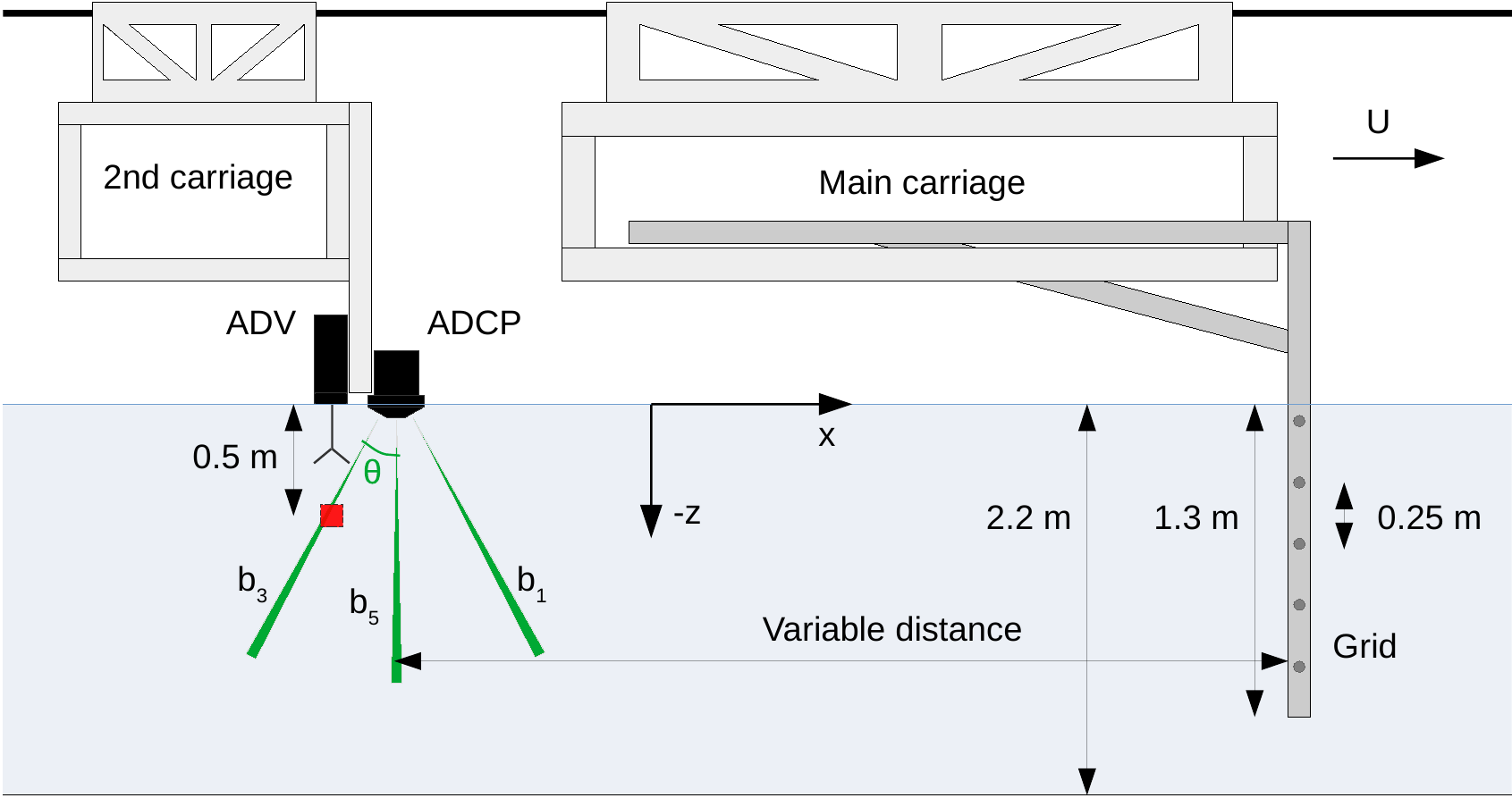}
    \caption{\label{fig:sketch} Schematics of the towing setup, here illustrated with the large grid. Both the ADCP and the ADV are illustrated (here mounted to the second carriage), but the two instruments were not deployed simultaneously. The red square indicates the position of the ADV measurement volume.} 
  \end{center}
\end{figure}

The experiments were conducted in a 50~m long, 3~m wide and 2.2~m deep towing and wave tank in the MarinLab hydrodynamic laboratory at the Western Norway University of Applied Sciences. A coordinate system was defined with the $(x,y,z)$-axis to be aligned in the axial, transverse and upward directions of the tank, respectively, with $z=0$ at the calm water surface. A computer-controlled carriage was towed along rails with a wire, and a second carriage could be coupled to the wire at any desirable distance behind the main carriage. The grids were hanged from a mounting frame fixed to the front of the main carriage, which was manufactured from 100$\times$50~mm sections EN~1.431 stainless steel, such that it was aligned with the $yz$-plane. Two regular biplane grids were used; a large grid with mesh size $M=0.25$~m and bar diameter $d=5$~cm and a small grid with $M=0.1$~m and $d=2$~cm. The grids were manufactured from 6060 aluminium circular-section tubes. The tubes were welded into separate aluminium frames of 2.5~mm thickness, and tube ends left open to allow flooding of the grid. Both grids had a solidity coefficient $\beta = 2d/M-(d/M)^2=0.36$, which is similar to the grids used in e.g. \cite{comte1971simple} and \cite{murzyn2005experimental}. The grids were located in the tank center and spanned 1.4~m in width and 1.3~m in depth. Images and a schematic of the grid towing setup are shown in Figs.~\ref{fig:towing_config} and \ref{fig:sketch}, respectively.  

The instruments were mounted at $x=$~1.5, 2.5 and 5~m behind the grid, either to the main carriage if $x<$~2~m or to the second carriage otherwise. The ADCP test matrix is listed in Table~\ref{Table:runs}. The ADV test matrix was identical, except that the instrument was mounted at $x=$~0.5, 2.5 and 5~m behind the grid when the large grid was applied. Separate runs were performed with each instrument. The ADCP was mounted with the transducers submerged a couple of centimeters below the surface with the horizontal components of $\mathbf{b_1-b_4}$ pointing in the $x$, $y$, $-x$ and $-y$- direction, respectively, and $\mathbf{b_5}$ pointing downwards. The ADV beams were aligned with the tank axes and the measurement volume was located at $z=-0.5$~m. The carriage was accelerated at 0.5~m/s$^2$ until a constant towing speed of $U=$~0.2 or 0.4~m/s (also used in \cite{liu1995energetics}) was reached. A safety distance to the wave maker at the one end and the damping beach at the other end of the tank had to be maintained. Therefore, the total towing length was 31-34~m, depending on the distance between the grid and the instruments. 

\begin{table}[h]
\centering % used for centering table
\begin{tabular}{c c c c c}  % centered columns (6 columns)
\toprule
\multirow{2}{*}{x [m]} & \multicolumn{2}{c}{M = 0.25 m} & \multicolumn{2}{c}{M = 0.1 m}  \\[0.5ex]
\cmidrule(lr){2-3} 
\cmidrule(lr){4-5} 
{} & U = 0.2 m/s & U = 0.4 m/s & U = 0.2 m/s & U = 0.4 m/s \\ [0.5ex]% inserts table
\midrule
1.5 & 97 & 95 & 88 & 65 \\[0.5ex]
2.5 & 93 & 95 & 78 & 85 \\[0.5ex]
5 & 85 & 90 & 88 & 90 \\[0.5ex]
\bottomrule
\
\end{tabular}
\caption[Grid turbulence test matrix]{Test matrix with the percentage of cells within $\mathbf{b_5}$ that satisfied the ADCP quality criterion (beam correlation greater than 50\%,  more than 90\% of the time).}
\label{Table:runs} % is used to refer this table in the text
\end{table}  

%The carriage was accelerated at 0.5~m/s$^2$ until a constant towing speed of $U=$~0.2 or 0.4~m/s (also used in \cite{liu1995energetics}) was reached. The mean and start/stop movement of the grid set up a seiche motion in the tank that was damped out after a couple of minutes. To ensure that the residual water motion was sufficiently damped out, new runs were initiated seven minutes after the carriage had been towed back to the starting position. A safety distance to the wave maker at the one end and the damping beach at the other end of the tank had to be maintained. Therefore, the total towing length was 31-34~m, depending on the distance between the grid and the instruments.  

The ADCP range was set to 1.1~m on all beams, including a blanking distance (where no measurements were performed) of 0.1~m close to the instrument head in order to avoid transducer ringing. The beam correlation, which is a data quality indicator described in Section~\ref{sec:processing}, was very sensitive to the instrument range, probably due to acoustic reflections in the tank. Several ranges were tested before adequate beam correlations were obtained. All the cells were located within the downstream grid area when the selected range was applied. Additionally, the ADCP beams did not reach the tank bottom and walls, hence sidelobe interference was avoided. The cell size was set to 2.5 and 5~cm for $U=$~0.2 and 0.4~m/s, which corresponded to 40 and 20 cells, respectively. The sampling frequency was set to the maximum possible value of 8 and 200~Hz for the ADCP and ADV, respectively. The tank was seeded with 10~$\mu$m spherical glass particles, and almost 5~kg of seeding particles was necessary to obtain satisfactory backscattering. The seeding particles were well distributed in the tank after a couple of initial runs with the grid. The difference between no seeding and seeding can be seen in Figs.~\ref{fig:towing_config_a} and \ref{fig:towing_config_b}, respectively.

\section{Data processing} \label{sec:processing}

From the raw data time series, data from the portion of the record that contained steady towing were carefully selected, i.e. data unaffected by acceleration and deceleration of the carriages. The rest of the time series was discarded. One run of steady towing lasted around 60-160~s, depending on the distance between the grid and the instruments and the towing speed, which means that the number of data points were in the order of $10^{3}$ and $10^{4}$ per run for the ADCP and the ADV, respectively. Due to the small amount of data points acquired with the ADCP, time series from several repetitions, typically two and four for towing speeds of 0.2 and 0.4~m/s, respectively, were concatenated to increase the number of independent data points. Only time series from single ADV runs were used, as these contained much more data points. For the ADCP, some TKE frequency spectra were estimated from longer data point segments and they were almost identical to those estimated from shorter segments, which indicates statistical convergence.   

\begin{figure}
  \begin{center}
    \includegraphics[width=.7\textwidth]{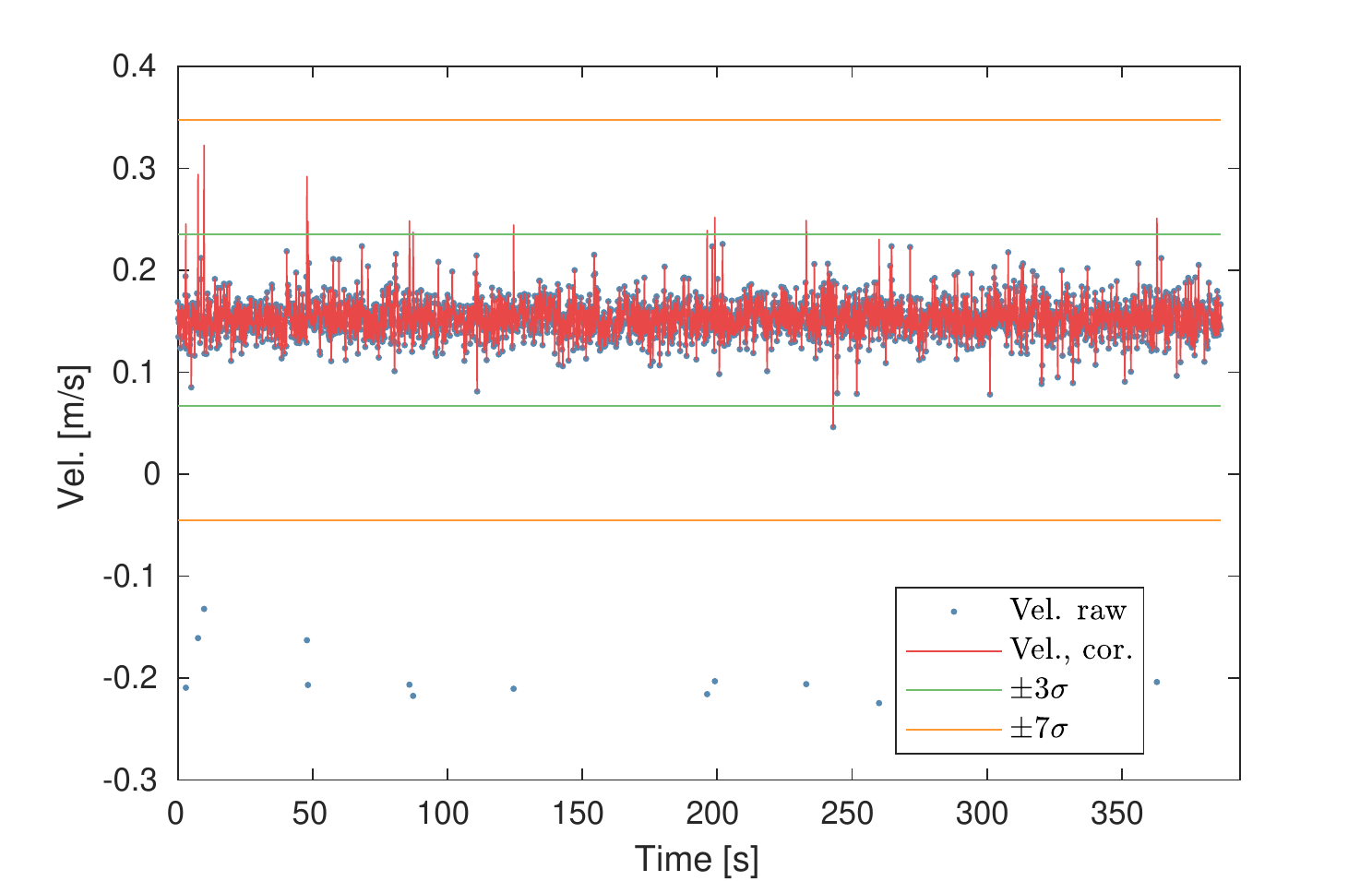}
    \caption[ADCP quality control]{\label{fig:QC} Quality control of a concatenated time series of steady towing at $U=$~0.4~m/s. Raw data are corrected for ambiguity wrapping when the velocity exceeds $\pm7\sigma$. Spikes are identified when the corrected velocity exceeds $\pm3\sigma$.}
  \end{center}
\end{figure}

The concatenated time series were quality controlled by inspecting the beam correlation, which should exceed 50\% for the ADCP and 70\% for the ADV per manufacturer recommendations. Data points with lower correlation than the recommended values were flagged, and time series containing more than 10\% low-correlation data were discarded. This was the case for some ADCP cells, typically far from the instrument transducer (an overview is given in Table~\ref{Table:runs} for $\mathbf{b_5}$), but all ADV time series contained satisfactory beam correlations. The ambiguity velocity $U_{amb}$ was approximately 0.25~m/s for the ADCP, and ambiguity wrapping occurred in some situations in the axially directed beams ($b_1$ and $b_3$). Data points located $\pm7\sigma$ off the sample mean, where $\sigma$ is the sample standard deviation, were identified as artifacts due to ambiguity wrapping and were unwrapped $\mp 2U_{amb}$ accordingly \citep{shcherbina2018observing}. Figure~\ref{fig:QC} shows an example of ambiguity-wrapped raw data (blue points) which are corrected (red line). 

Some spikes occurred in the time series, resulting typically from unphysical data such as low correlation data or acoustic contamination specific to the laboratory, but also from natural extreme values. Spikes were identified as data points located $\pm3\sigma$ off the sample mean \citep{nystrom2007evaluation}, indicated in Fig~\ref{fig:QC}. In calculations of statistical parameters, such as variance, the spikes were discarded. However, in spectral analysis, continuous time series are required. \cite{nystrom2007evaluation} include the spikes in the estimation of spectra. In the present study, data points that exceeded $\pm3\sigma$ off the sample mean were simply "cut" to $\pm3\sigma$ off the sample mean, i.e. at the green lines in Fig.~\ref{fig:QC}. This strategy may reduce the effect of natural extreme values but the noise from unphysical data is mitigated. Some spectra were estimated with both strategies, and they were almost identical, except that the spectra from the "cut" time series were marginally less energetic, which may indicate that the noise level was reduced. 

%Different approaches for handling spikes in the spectral analysis are reported in the literature. For example, \cite{zippel2018turbulence} interpolate neighboring data points in case of low correlation which may lead to spikes. Both techniques will necessarily give some errors due to the random nature of turbulence in the former case, especially if the turbulent scales are comparable to the sampling frequency, and the limited data quality in the latter case, particularly if there are solid boundaries in the vicinity of the instrument.     

Statistical analysis was performed on the quality-controlled data. Although homogeneity between the instantaneous beam velocities was not obtained for the ADCP, homogeneity was assumed in the mean and the variance of the signal \citep{stacey1999measurements}. The mean axial fluid velocity relative to the instrument $\langle u \rangle$, where the angled brackets denote time averaging over the whole time series, was obtained directly from the horizontal component of the ADV and from $\langle u \rangle=(\langle b_1 \rangle-\langle b_3 \rangle)/2\cos\theta$ of the ADCP \citep{stacey1999measurements}. The fluctuating velocity component in any direction $u'_i=u_i- \langle u_i \rangle$ was used in the turbulence analysis. Data that were flagged as spikes or with correlation less than the recommended values were discarded before calculating the velocity variance $\langle u'^{2}_{i} \rangle$. The component of the TKE in the vertical direction $TK_z$ was defined as $TK_z = \langle w'^{2} \rangle$. Since the vertical velocity was directly measured with the ADCP ($b_5$), no beam transformation was required to estimate $TK_z$. Following \cite{dewey2007reynolds}, the total TKE density $TK$ was calculated as

\begin{equation}
\label{eq:TKE_ADV}
TK_{ADV} = \rho_w\frac{\langle u'^2 \rangle + \langle v'^2 \rangle + \langle w'^2 \rangle}{2},
\end{equation}

\begin{equation}
\label{eq:TKE_ADCP}
TK_{ADCP} = \rho_w\frac{\langle b'^{2}_{1}\rangle + \langle b'^{2}_{2}\rangle + \langle b'^{2}_{3}\rangle + \langle b'^{2}_{4}\rangle - 2(2 \cos^2 \theta - \sin^2 \theta) \langle b'^{2}_{5}\rangle}{4 \sin^2 \theta},
\end{equation}

\noindent for the ADV and the ADCP, respectively, where $\rho_w=$~1000~kg/m$^3$ is the water density. Equation~\ref{eq:TKE_ADCP} combines the variance estimates from the ADCP transducers according to vector algebra to estimate the Cartesian 3D variance components given in Eq.~\ref{eq:TKE_ADV}, with the assumption of homogeneity in variance over distances comparable to the horizontal separation of the bins \citep{dewey2007reynolds}.   

Turbulent kinetic energy frequency spectra $PSD_{w}(f)$ were estimated from the vertical fluctuating velocity component $w'$ with the Welch method \citep{earle1996nondirectional}, which means fast Fourier transformation and ensemble averaging of overlapping segments. Each time series was divided into segments of 1024 data points with 50\% overlap and a Hamming window was applied to each segment to reduce spectral leakage. The TKE frequency spectra represent the distribution of turbulent kinetic energy over the frequencies $0<f<f_N$, where $f_N$ is the Nyquist frequency, which was 4 and 100~Hz for the ADCP and the ADV, respectively. From Fig~\ref{fig:QC}, it is clear that $u'_{i}/U \ll 1$, which indicates that the advection of turbulence past the instrument is dominated by the mean flow and not by the circulation of eddies, and that the assumption of Taylor's hypothesis is valid. Therefore, the TKE frequency spectra should be proportional to $f^{-5/3}$ in the inertial subrange.    

Doppler noise often results in flat TKE frequency spectra, also known as the noise floor, towards the higher frequencies where the turbulent energy is low. From inspections of ADV data, it was observed that the noise floor was reached close to the Nyquist frequency, and the noise floor was found by averaging the 20 highest frequencies of the TKE spectra, which corresponds to frequencies in the range 96-100~Hz. Following \cite{thomson2012measurements}, the noise variance $n^2$ of the ADV was estimated by integrating the noise floor over the range of frequencies $0<f<f_N$, assuming white noise spectra. The Doppler noise was removed from the ADV velocity variance statistically \citep{lu1999using} by subtracting the noise variance, so that $\langle u'^{2}_{i} \rangle=var(u'_i)-n^2$. It was not clear whether the ADCP spectra, which are shown in Section~\ref{sec:results}, were obscured by noise. Hence, the turbulence properties obtained from the ADCP were not corrected for Doppler noise.

\section{Results and discussion} \label{sec:results} 

\begin{figure}
  \begin{center}
    \includegraphics[width=.95\textwidth]{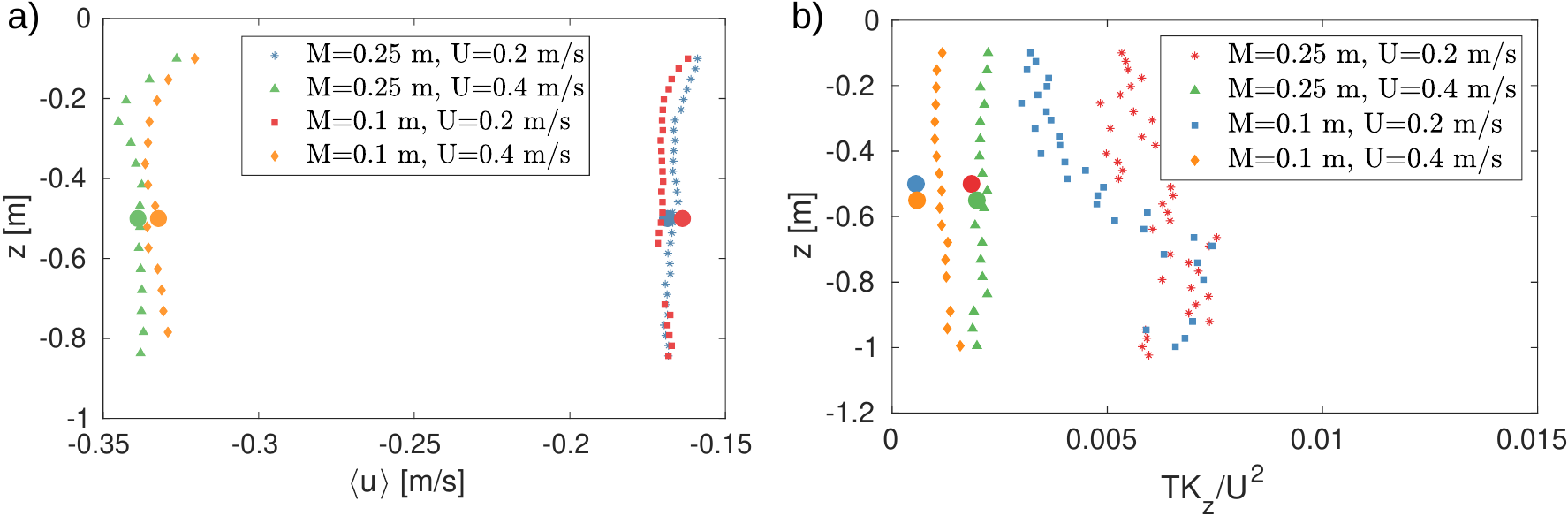}
    \caption[ADCP quality control]{\label{fig:velocity_profile} Vertical ADCP profiles at $x=2.5$~m of a) mean axial velocity and b) $TK_z/U^2$. The large dots are corresponding ADV data. Some of the ADV data points are slightly displaced in the vertical direction to increase the readability. Some data points are omitted due to low beam correlations.}
  \end{center}
\end{figure}

The ADCP was able to reproduce the mean velocity in the axial direction, as can be seen in Fig.~\ref{fig:velocity_profile}a). For the cells located at the same vertical level as the ADV measurement volume, the errors were less than 4\%. Some missing cells were discarded because the beam correlation criterion stated in Section~\ref{sec:processing} (beam correlation must be greater than 50\% more than 90\% of the time) was not fulfilled. The measured velocities were about 10\% lower than the towing speeds due to the grid-induced velocity deficit. The component of the TKE in the vertical direction $TK_z$, non-dimensionalized over the square of the towing speed $U^2$, is presented in Fig.~\ref{fig:velocity_profile}b). There was good agreement between the instruments for the high towing speed and the large grid, but the ADCP largely overestimated $TK_z/U^2$ in the other situations (by a factor of 2 for $U$ = 0.4~m/s and approximately 5 for $U$ = 0.2~m/s). A possible reason for this could be Doppler noise in the ADCP, which can cause a high bias in estimators related to TKE \citep{stacey1999measurements}. As stated in Section~\ref{sec:processing}, estimated variance from the ADCP was not corrected for Doppler noise in the post processing.      

\begin{figure}
  \begin{center}
    \includegraphics[width=.7\textwidth]{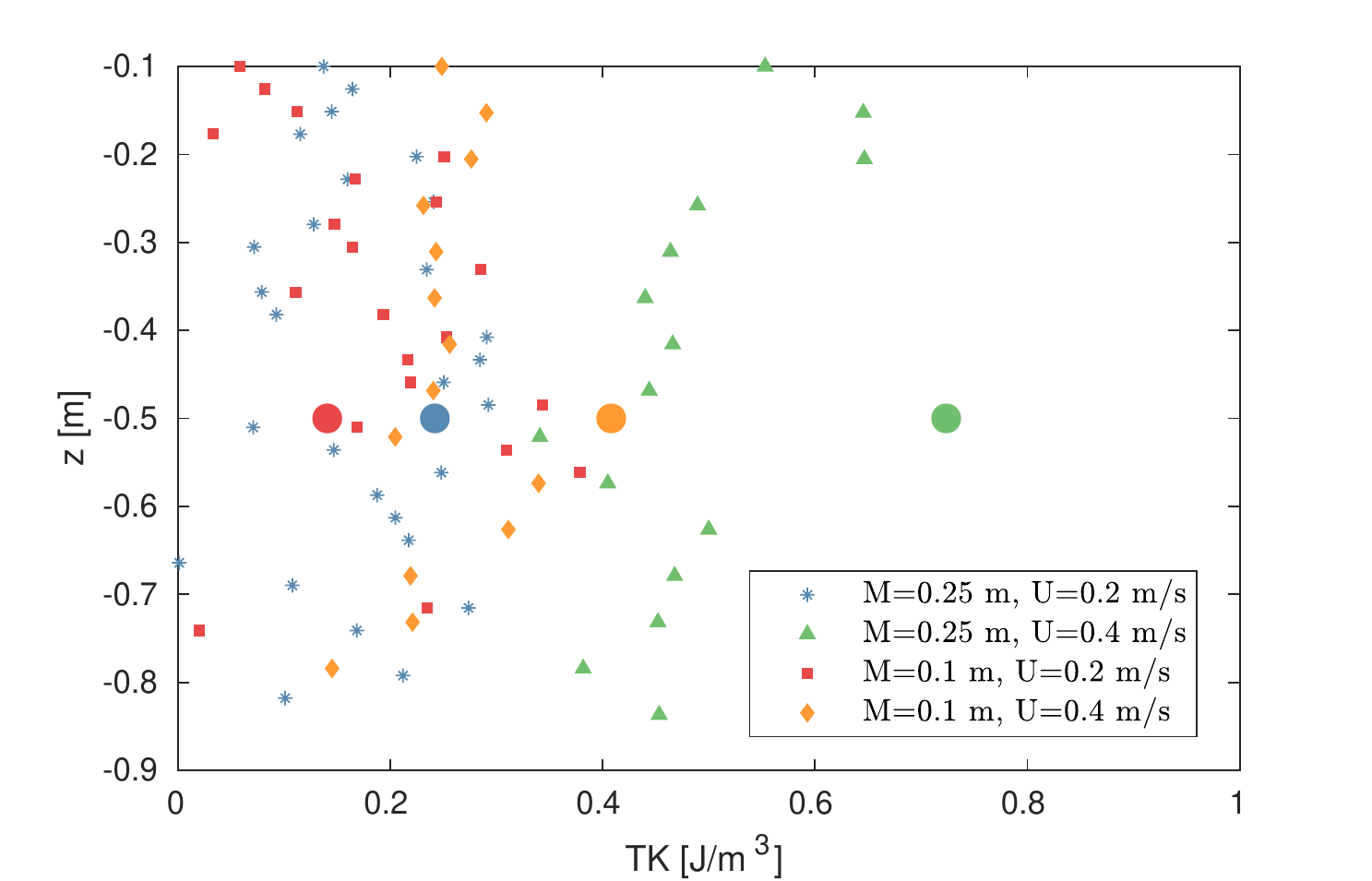}
    \caption[Turbulent kinetic energy profiles]{\label{fig:TKE_profile} Profiles of $TK$ estimated from the ADCP and single values from the ADV at $x=2.5$~m. See the figure title of Fig.~\ref{fig:velocity_profile} for further details.}
  \end{center}
\end{figure}

Although the ADCP overestimated $TK_z$ in the vertical direction, this was not the case for the total TKE densities $TK$ estimated from Eqs.~\ref{eq:TKE_ADV}-\ref{eq:TKE_ADCP}, presented in Fig.~\ref{fig:TKE_profile}. For the high towing speed, the values estimated from the ADCP were up to 35\% smaller than the equivalent ADV values. The instruments agreed fairly well for the low towing speed, although these estimates were quite scattered in the vertical profile. Doppler noise, which could vary between beams and cells, may have caused the deviations. Velocity components in all directions were used to estimate $TK$. It was observed from the ADV data that the grid-generated turbulence was not very isotropic. The ratio of $\langle w'^2 \rangle/\langle u'^2 \rangle$ was about 0.6, which is similar to the values obtained in another towing tank \citep{liu1995energetics}. In comparison, grid-generated turbulence in a wind tunnel could reach a ratio closer to unity, for example, \cite{comte1966use} reported the ratio~0.8. Complete isotropy is not attainable in grid-generated turbulence due to the spatial inhomogeneity that arise from decay of TKE in the downstream direction \citep{comte1966use}.       

\begin{figure}
  \begin{center}
    \includegraphics[width=.7\textwidth]{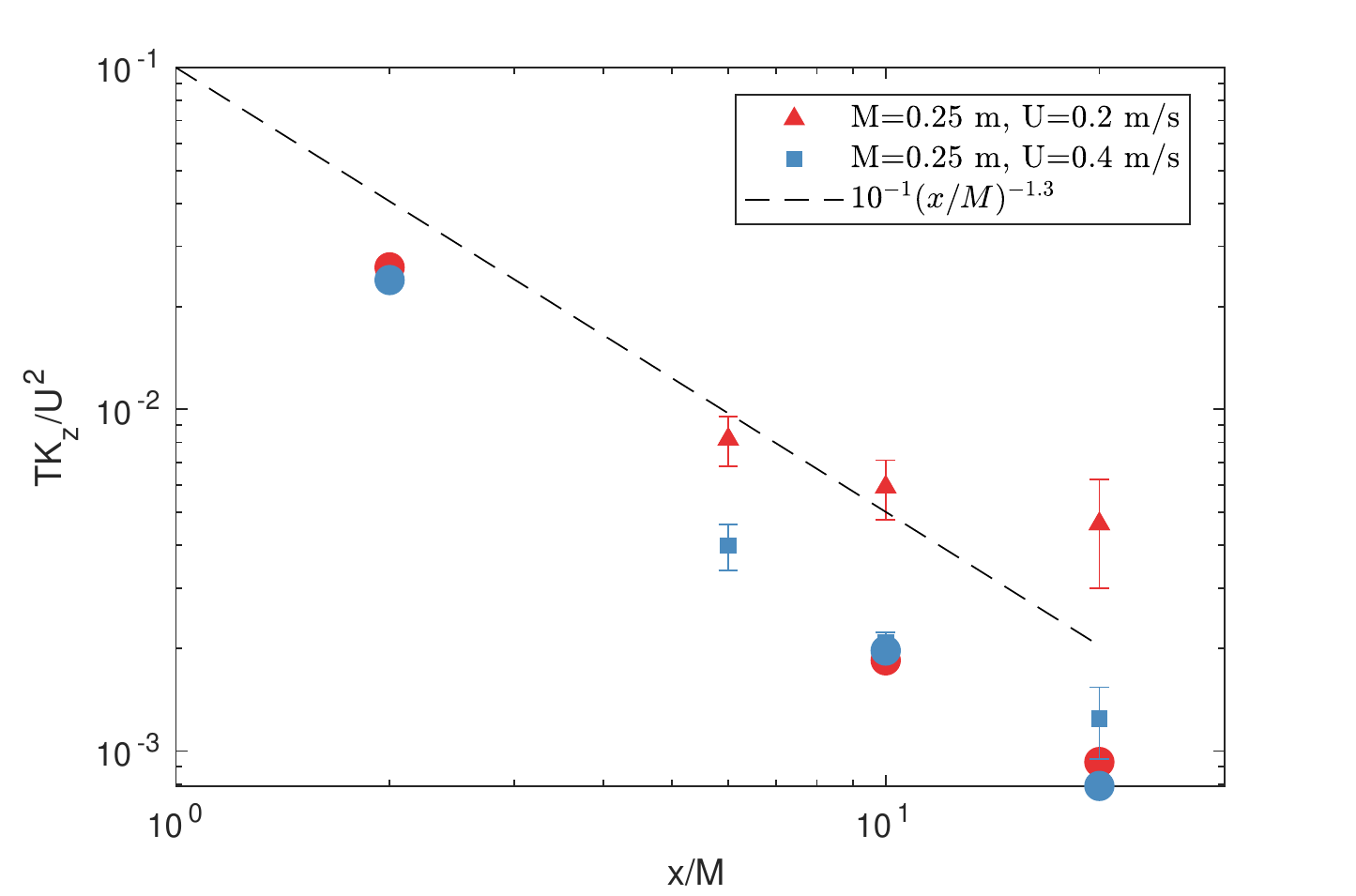}
    \caption[Turbulent decay downstream of a grid]{\label{fig:turbulence_decay} Decay of $TK_z/U^2$ as function of distance downstream of the grid for $M=0.25$~m. ADCP cells (11 in most cases) are averaged to obtain the data points (${\color{red}\triangle}$ and ${\color{blue}\square}$) and the error bars show two standard deviations of the sample of cells. The large dots are corresponding ADV data.}
  \end{center}
\end{figure}

Figure~\ref{fig:turbulence_decay} shows $TK_z/U^2$ versus downstream distance $x/M$ for the large grid. The squares and triangles indicate the mean values obtained by ensemble averaging the ADCP cells located close to the ADV measurement volume in terms of depth. Typically, 11 cells were used to estimate the mean value, provided that all the time series satisfied the beam correlation criterion. A decay in $TK_z$ can be observed, and the vertical components of the TKE decrease with increasing downstream distance approximately as $(x/M)^{-1.3}$, especially in the case of the ADV, which agrees with the findings of \cite{sirivat1983effect} and \cite{liu1995energetics}. The ADCP data from the high towing speed approximately follow the same slope in the log-log plot, while the low towing speed data exhibit a weaker decay, which is in accordance with the mismatch observed between the ADV and the ADCP in $TK_z/U^2$ for the low towing speeds in Fig.~\ref{fig:velocity_profile}b). The low towing speed data also have large error bars for increasing downstream distance. The error bars represent the spread of the data obtained from the different cells and show two standard deviations of the sample of (11) cells. We do not attempt to justify the existence of a power law, as the data hardly span over one decade in distance, and only three data points are considered. %More measurements at larger distances would have been advantageous to evaluate the decay of turbulence.       

\begin{figure*}
        \centering
        \begin{subfigure}[b]{0.475\textwidth}
            \centering
            \includegraphics[width=\textwidth]{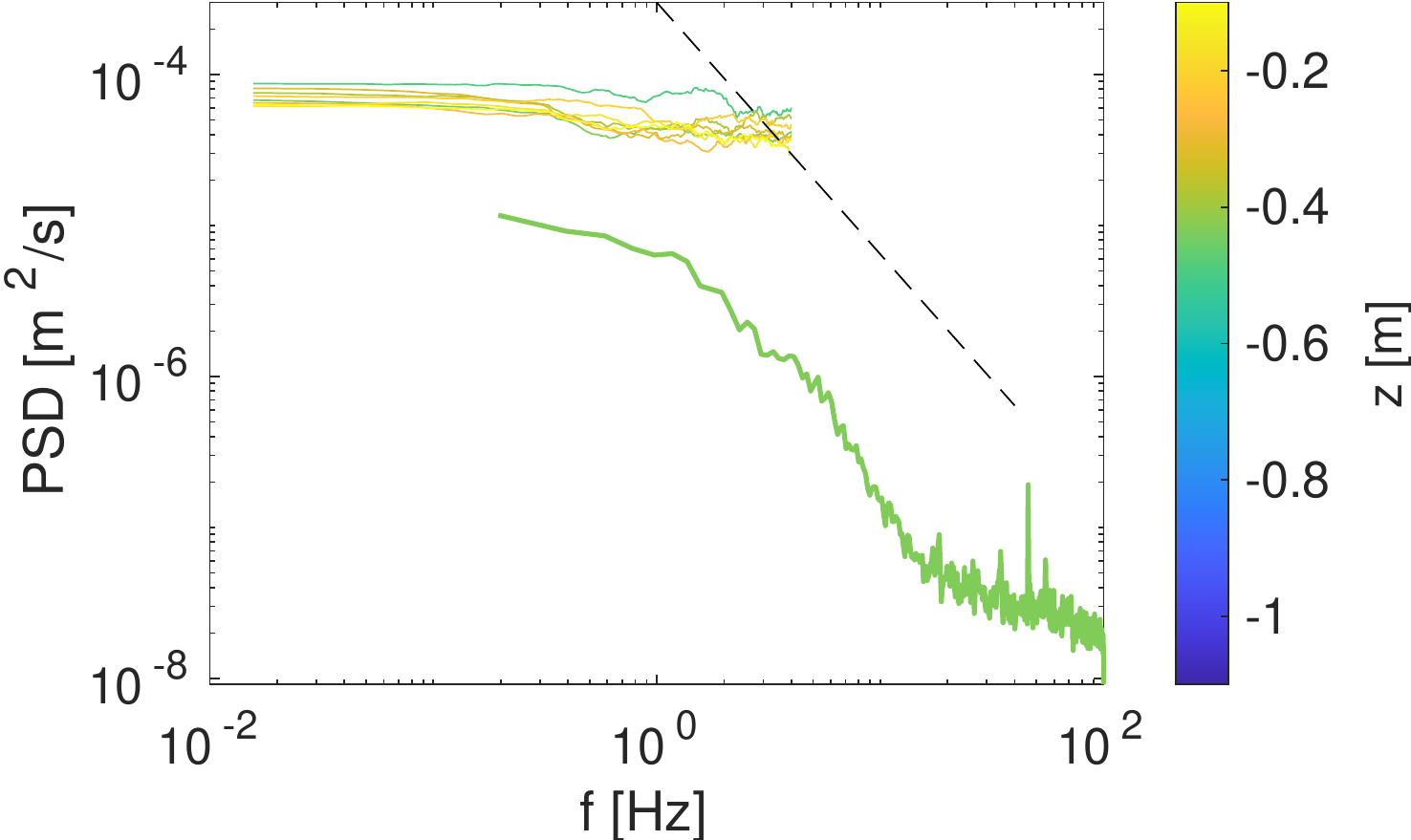}
            \caption[Network2]%
            {{\small $M$ = 0.1~m, $U$ = 0.2~m/s}}    
            \label{fig:mean and std of net14}
        \end{subfigure}
        \hfill
        \begin{subfigure}[b]{0.475\textwidth}  
            \centering 
            \includegraphics[width=\textwidth]{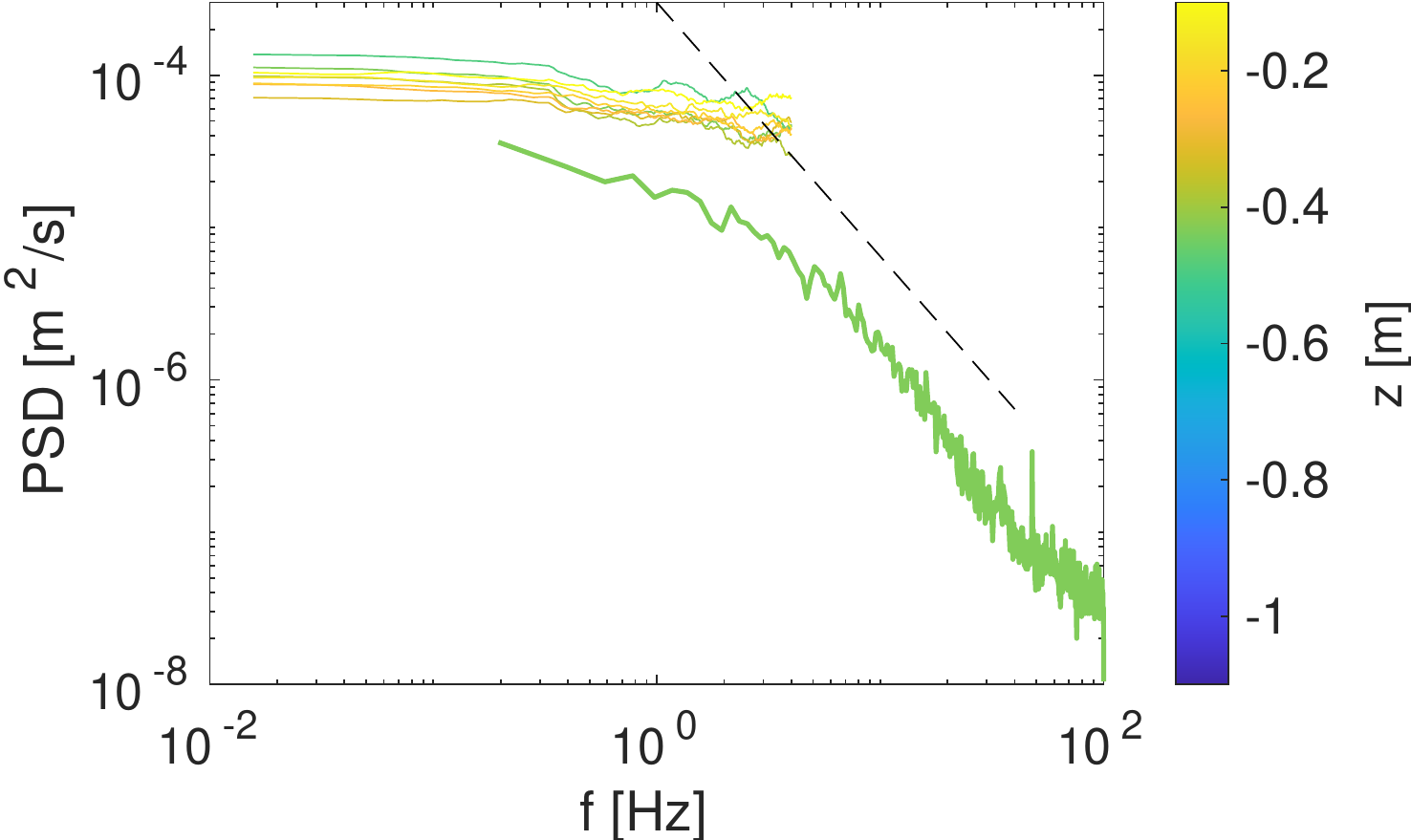}
            \caption[]%
            {{\small $M$ = 0.1~m, $U$ = 0.4~m/s}}    
            \label{fig:mean and std of net24}
        \end{subfigure}
        \vskip\baselineskip
        \begin{subfigure}[b]{0.475\textwidth}   
            \centering 
            \includegraphics[width=\textwidth]{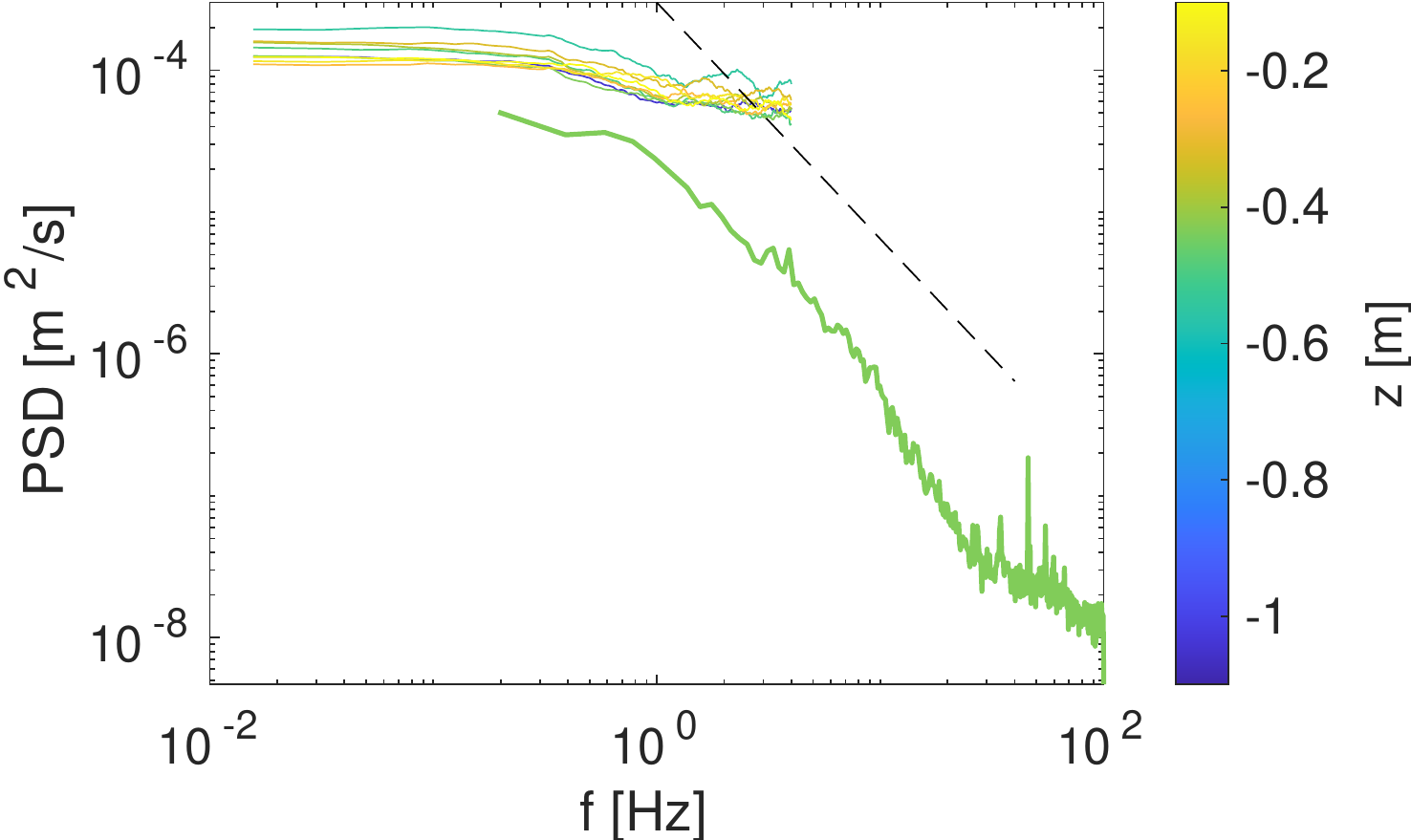}
            \caption[]%
            {{\small $M$ = 0.25~m, $U$ = 0.2~m/s}}    
            \label{fig:mean and std of net34}
        \end{subfigure}
        \hfill
        \begin{subfigure}[b]{0.475\textwidth}   
            \centering 
            \includegraphics[width=\textwidth]{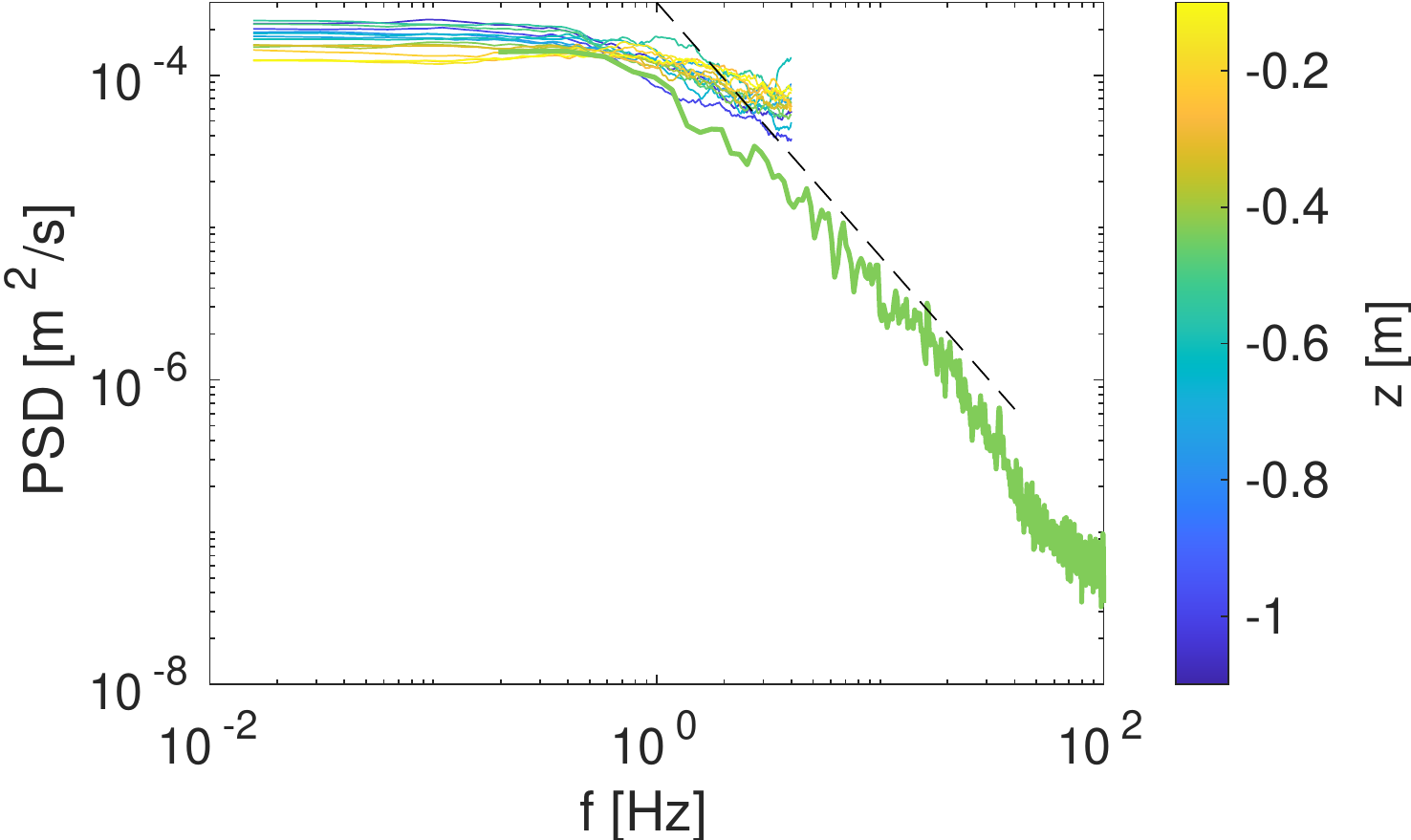}
            \caption[]%
            {{\small $M$ = 0.25~m, $U$ = 0.4~m/s}}    
            \label{fig:mean and std of net44}
        \end{subfigure}
        \caption[Power spectral densities]
        {\small Turbulent kinetic energy spectra from the vertical velocity component at $x=2.5$~m, distributed over the vertical profile. The thick, green line is ADV data, and the dashed line is proportional to the theoretical -5/3 power law in the inertial subrange.} 
        \label{fig:spectra}
    \end{figure*}

Turbulent kinetic energy spectra estimated from the vertical fluctuating velocity components of the two instruments are presented in Fig.~\ref{fig:spectra}. Spectra from the ADCP at 10 positions evenly distributed over the vertical profile are included, provided that the time series satisfied the beam correlation quality criterion. In general, it can be observed that the TKE level was higher with increasing towing speed and grid mesh size. The ADV spectra were proportional to the theoretical -5/3 power law over a wide range of frequencies, meaning that the instrument resolved the inertial subrange of turbulence. The ADV noise floor was reached well above 10~Hz and the noise level was $\sim10^{-8}~\mathrm{m}^{2}\mathrm{s}^{-1}$. However, the ADCP was not able to resolve the inertial subrange. The ADCP spectra appear to be quite flat around $10^{-4}~\mathrm{m}^{2}\mathrm{s}^{-1}$, which is consistent with the noise level of the Nortek Signature reported by \cite{guerra2017turbulence} from tidal channel measurements.  

There are several possible explanations for the ADCP's failing ability to resolve the inertial subrange of turbulence. First, Doppler noise seems to prevent the instrument from detecting turbulent energy below $\sim10^{-4}~\mathrm{m}^{2}\mathrm{s}^{-1}$. This can be seen when the TKE starts to decrease with increasing frequency in accordance with the ADV spectra around 0.5~Hz, but then remain approximately constant. In other words, the TKE level at the scales associated with the inertial subrange in the flow may be too low to be detected by the instrument. On the other hand, it is not clear that the spectra flatten out in all situations. For example, in Fig.~\ref{fig:spectra}d), the ADV and ADCP spectra have approximately the same slope until the Nyquist frequency of the ADCP is reached, which makes it difficult to conclude that the noise floor obscures the spectra. 

Another possible explanation for the relatively flat spectra is the fact that the ADCP measurement volume may not be adequately small to resolve the scales $l_{IS}$ associated with the inertial subrange. In fact, water flume experiments on grid-generated turbulence suggest that the integral scale of turbulence in the axial direction $L_x$ is approximately $M/2$ at the downstream position $x/M\approx20$ and slowly increasing with $x/M$ \citep{murzyn2005experimental}. The same findings are reported in the wind tunnel experiments of \cite{comte1971simple}, who also found that the integral length scales $L_y$ and $L_z$ in the transverse and vertical directions, respectively, were about half the size of $L_x$. These observations may indicate that $L_z \sim M/4$ at the downstream distances investigated in the present study. When the large grid is considered, the ADCP cell size was $M/10$ and $M/5$ for $U$~=~0.2~m/s and 0.4~m/s, respectively, which means that probably only the largest turbulent eddies were resolved by the instrument. The cell size to grid mesh ratio was even larger for the small grid.  

\section{Conclusions} \label{sec:conclusion}

In this work, the ability of the Nortek Signature1000 five beam ADCP operated in high resolution mode to measure fine-scale turbulence was investigated. The provided experimental validation of the effective cutoff frequency and turbulent eddy size, indicates under which flow conditions the instrument can resolve turbulence. The research was carried out in a large towing tank where turbulence was generated from a towed grid. An ADV was used as reference and the two instruments were in turn towed at a fixed distance behind the grid. Turbulence was described through the statistical parameters; variance, TKE and TKE spectra. The mean flow measured with the ADCP was accurate within 4\% of the ADV. Variance in the vertical direction $TK_z$ was reasonably well resolved by the ADCP when the large grid was applied with the high towing speed, but largely overestimated by a factor of 2-5 for the low towing speed and small grid. This deviation was attributed to either Doppler noise, too small grid-generated eddies with respect to the sample volume/frequency or a combination of these. Better agreement was observed for total TKE density $TK$, although the data were quite scattered along the vertical profile. 

On the one hand, TKE and velocity variance are predominantly defined by the larger energetic eddies and were probably therefore reasonably well estimated by the ADCP in some grid size and towing speed combinations. On the other hand, spectra represent the distribution of TKE over a range of scales, and the minimum resolvable scale and frequency is limited by half the cell size and sampling frequency, respectively. Therefore, the ADCP was probably only able to measure the largest scales of the flow, and the inertial subrange was not visible in the spectra. Doppler noise may also have obscured smaller eddies around the Nyquist frequency. The spectral noise level of the Nortek Signature found in the towing tank was consistent with field experiments in tidal channels where the integral turbulent scales were much larger \citep{guerra2017turbulence}. The present results indicate that the instrument is not suitable for grid turbulence measurements of this scale with regards to spectral properties. The largest turbulent scales should probably be at least an order of magnitude larger than the cell size to be able to resolve the inertial subrange, which may be difficult to obtain with grid generated turbulence in a laboratory. 

%Acoustic profilers operated in the high-resolution mode suffer from a relatively small velocity range, meaning that ambiguity errors may occur if the flow velocity is high. Occasional ambiguity wrapping was observed in the axially directed beams for the high towing speed, but these data points were easily identified and unwrapped since the mean flow was constant. However, if the mean flow is not constant, for example oscillating with periods $\sim 10$~s, which was the case in Paper~V in this thesis, ambiguity wrapping could be hard to identify. The data presented in Paper~V may contain some ambiguity errors in the axially directed beams, but this is challenging to determine with certainty due to the fluctuating nature of the flow in combination with the noise, which was prominent in some cases. Therefore, spectral turbulent properties were only obtained from the vertical beam in Paper~V, and the problem of ambiguity wrapping was avoided in the spectra.    

%If the experiments are to be repeated, the grid and mesh size should be increased as much as possible. The number of varied parameters could be increased in future experiments. It would for example add value to the study if the ADV depth is changed to allow for comparison along the entire ADCP profile. It may also be interesting to investigate the sensitivity of the cell size with respect to the results. Also, more repetitions with the same parameters should be ran with the ADCP to strengthen the validity of the results. 

\section*{Acknowledgement}

The authors are grateful to Yiyi Whitchelo, Gloria Stenfelt and Jan Bartl for their assistance during the laboratory work. Funding for the experiment was provided by the Research Council of Norway under the PETROMAKS2 scheme (project DOFI, Grant number $28062$). The data are available from the corresponding author upon request.

\begin{footnotesize}
\bibliographystyle{agsm} % agsm was jfm
\bibliography{Bergen_arxiv.bib}
\end{footnotesize}

\end{document}